\documentclass[12pt]{article}
\usepackage{epsf}
\usepackage{amsmath}

\usepackage{graphics}
\usepackage{cite}

\begin{titlepage}

\begin{document}

\hfill July 2021\\

\begin{center}

{\bf \Large Additional Baryons and Mesons\\
}
\vspace{2.5cm}
{\bf Paul H. Frampton\footnote{ paul.h.frampton@gmail.com} \\  }
\vspace{0.5cm}
{\it Dipartimento di Matematica e Fisica "Ennio De Giorgi",\\ 
Universit\`{a} del Salento and INFN-Lecce,\\ Via Arnesano, 73100 Lecce, Italy
}

\vspace{0.7in}

\Large

\begin{abstract}
\noindent
In a particle theory model whose most readily discovered new particle is the $\sim 1$TeV bilepton resonance in same-sign leptons, currently being sought at CERN's  LHC, there exist three quarks ${\cal D, S, T}$ 
which will be bound by QCD into baryons and mesons. We consider the decays of these additional baryons and mesons whose detailed experimental study will be beyond
the reach of the 14 TeV CERN collider and accessible only at an O(100 TeV) collider.
\end{abstract}

\end{center}
\end{titlepage}

\noindent
An important question in high-energy physics is what will be the first new particle to be
discovered beyond the standard model. In the Letter, we discuss a possible answer to this
question and how our answer could impact on particle phenomenology beyond the TeV energy
scale.

\bigskip

\noindent
A particle theory model which explains the occurrence of 
three quark-lepton families in the standard model \cite{Glashow,Weinberg,Salam} is the bilepton model. 
For the minimal 331-model with no additional leptons, 
as proposed in \cite{PHF1992}, this involves, {\it inter alia}, doubly-charged gauge bosons
$Y^{\pm\pm}$ with 
lepton number $|L|=2$ which decay into same-sign leptons and are being
sought at the LHC. Some of the relevant LHC phenomenology is discussed
in \cite{CCCF1,CCCF2}. A refined mass estimate derived in \cite{CF}
is $M(Y^{\pm\pm}) = (1.29 \pm 0.06)$ TeV where {\it faute de mieux} it was assumed that the
symmetry breaking of $SU(3)_L$ is closely similar to that of $SU(2)_L$. It will be pleasing
if the physical mass is consistent with this.

\bigskip

\noindent
Because the quarks are in triplets and anti-triplets of $SU(3)_L$, rather than only in doublets of $SU(2)_L$ as in the
standard model, there is necessarily an additional quark in each family. In the first and second families they are the
${\cal D}$ and ${\cal S}$ respectively, both with charge $Q=-4/3$ and lepton number $L=+2$. In the third family 
is the ${\cal T}$
with charge $Q=+5/3$ and lepton number $L=-2$. All the three TeV scale quarks are colour triplets with
spin-$\frac{1}{2}$ and baryon number $B=\frac{1}{3}$. Their masses
are yet to be measured but may be expected to be below the ceiling of $4.1 TeV$ which is the upper limit
for symmetry breaking of $SU(3)_L$ and probably above $1 TeV$. By analogy with the
known quarks, one might expect $M({\cal T}) > M({\cal S}) > M({\cal D})$, although 
without experimental data this is
conjecture.

\bigskip

\noindent
The heavy quarks and antiquarks will be bound to light quarks and antiquarks, and to each other, to form an interesting spectroscopy of
mesons and baryons. Let us first display, in Tables 1, 2 the TeV mesons, then in Tables 3,4,5 the TeV baryons. The charge conjugate
states are equally expected, and will reverse the signs of $Q$ and $L$.

\bigskip

\begin{table}[h]
\caption{TeV mesons ${\cal Q}\bar{q}$}
\begin{center}
\begin{tabular}{||c|c||c|c||}
\hline
\hline
${\cal Q}$ & $\bar{q}$ & Q & L \\
\hline
\hline
&&& \\
${\cal D/S}$ & $\bar{u}$ etc. & -2 & +2 \\
 ${\cal D/S}$ & $\bar{d}$ etc. & -1 & +2 \\
 ${\cal T}$ & $\bar{u}$ etc. & +1 & -2 \\
  ${\cal T}$ & $\bar{d}$ etc.& +2 & -2 \\
  &&& \\
\hline
\hline
\end{tabular}
\end{center}
\label{mesonsQq}
\end{table}

\newpage

\begin{table}[t]
\caption{TeV mesons ${\cal Q} \bar{{\cal Q}}$}
\begin{center}
\begin{tabular}{||c|c||c|c||}
\hline
\hline
& & & \\
 ${\cal Q}$ & $\bar{{\cal Q}}$ & Q & L \\
\hline
&&&\\
${\cal D/S}$ & ${\cal \bar{D}/\bar{S}}$ & 0 & 0 \\
${\cal D/S}$ & $\bar{{\cal T}}$ &  -3 & +4 \\
${\cal T}$ & $\bar{T}$ & 0  & 0 \\
&&& \\
\hline
\hline
\end{tabular}
\end{center}
\label{mesonsQq}
\end{table}

\bigskip

\begin{table}[h]
\caption{TeV baryons ${\cal Q}qq$}
\begin{center}
\begin{tabular}{||c|c||c|c||}
\hline
\hline
&&& \\
 ${\cal Q}$ & qq & Q & L \\
\hline
&&&\\
${\cal D/S}$ & dd etc. & -2 & +2 \\
${\cal D/S}$ & ud etc. & -1 & +2 \\
${\cal D/S}$ & uu etc. & 0  & +2 \\
${\cal T}$ & dd etc. & +1 & -2 \\
${\cal T}$ & ud etc. & +2 & -2 \\
${\cal T}$ & uu etc, & +3 & -2 \\ 
&&& \\
\hline
\hline
\end{tabular}
\end{center}
\label{mesonsQq}
\end{table}

\newpage

\begin{table}[t]
\caption{TeV baryons ${\cal Q}{\cal Q}q$}
\begin{center}
\begin{tabular}{||c|c||c|c||}
\hline
\hline
& & & \\
 ${\cal Q}{\cal Q} $ & $q$ & Q & L \\
\hline
&&&\\
${\cal (D/S)(D/S)}$ & d etc. & -3 & +4 \\
${\cal (D/S)(D/S)}$ & u etc.& -2  & +4 \\
${\cal (D/S)T}$ & d etc. & 0 & 0 \\
${\cal (D/S)T}$ & u etc.&+1 & 0  \\
${\cal TT}$ & d etc. & +3 & -4 \\
${\cal TT}$ & u etc. & +4 & -4 \\ 
&&& \\
\hline
\hline
\end{tabular}
\end{center}
\label{mesonsQq}
\end{table}

\begin{table}[h]
\caption{TeV baryons ${\cal Q}{\cal Q}{\cal Q}$}
\begin{center}
\begin{tabular}{||c||c|c||}
\hline
\hline
& &  \\
 ${\cal Q}{\cal Q}{\cal Q}$ & Q & L \\
\hline
&&\\
${\cal (D/S)(D/S)(D/S)}$ & -4 & +6 \\
${\cal (D/S)(D/S)T}$ & -1  & +2 \\
${\cal (D/S)TT}$ & +2 & -2 \\
${\cal TTT}$ & +5 & -6  \\ 
&& \\
\hline
\hline
\end{tabular}
\end{center}
\label{mesonsQq}
\end{table}

\noindent
Although the ${\cal Q}$ masses are unknown, it may be reasonable first
to make a preliminary discussion of these states by assuming that
\begin{equation}
M({\cal T})> M({\cal S}) + 2M_t > M({\cal D}) +4M_t
\label{simplification}
\end{equation}
where $M_t$ is the top quark mass so that the lightest of the TeV baryons and mesons
are those containing just one ${\cal D}$ quark or one ${\cal \bar{D}}$ antiquark. The next lightest are
the TeV baryons and mesons containing just one ${\cal S}$ quark or one $\bar{{\cal S}}$ antiquark.

\newpage

\noindent
We begin by discussing the decay modes of the ${\cal D}\bar{q}$
mesons in Table 1, focusing on final states from the first family.
The decays of ${\cal D}$ include, taking care of $L$ conservation,
\begin{eqnarray}
{\cal D} &\rightarrow& d + Y^- \nonumber \\
& \rightarrow & d + (e^- + \nu_e) \nonumber \\
& \rightarrow & d + (\mu^- + \nu_{\mu}) \nonumber \\
& \rightarrow & d + (\tau^- + \nu_{\tau}) \nonumber \\
\label{calDdecay}
\end{eqnarray}
which implies that decays of the $({\cal D} \bar{u})$ meson include
\begin{eqnarray}
({\cal D} \bar{u}) & \rightarrow & \pi^- + (e^- + \nu_e) \nonumber \\
& \rightarrow & \pi^- + (\mu^- + \nu_{\mu}) \nonumber \\
& \rightarrow &  \pi^- + (\tau^- + \nu_{\tau}) \nonumber \\
\label{calDudecay}
\end{eqnarray}
and variants thereof where $\pi^-$ is replaced by any other non-strange
negatively charged meson. The $d$ in Eq.(\ref{calDdecay}) can be
replaced by $s$ or $b$ which subsequently decay.

\bigskip

\noindent
An alternative to Eq.(\ref{calDdecay}) is
\begin{eqnarray}
{\cal D} &\rightarrow& u + Y^{- -} \nonumber \\
& \rightarrow & u + (e^- + e^-) \nonumber \\
& \rightarrow & u + (\mu^- + \mu^-) \nonumber \\
& \rightarrow & u + (\tau^- + \tau^-) \nonumber \\
\label{calDdecay2}
\end{eqnarray}
which implies additional decay modes of the $({\cal D} \bar{u})$ meson which include
\begin{eqnarray}
({\cal D} \bar{u}) & \rightarrow & \pi^0 + (e^- + e^-) \nonumber \\
& \rightarrow & \pi^0 + (\mu^- +\mu^-) \nonumber \\
& \rightarrow &  \pi^0 + (\tau^- + \tau^-) \nonumber \\
\label{calDudecay2}
\end{eqnarray}
and variants obtained by flavour replacements.  Eqs.(\ref{calDudecay}) and (\ref{calDudecay2}),
and their generalisations to other flavours,
suffice to illustrate the richness of $({\cal D}\bar{u})$ decays.

\newpage

\noindent
Turning to the meson ${\cal D} \bar{d}$, we can use
Eq.(\ref{calDdecay}) to identify amongst its possible decays
\begin{eqnarray}
({\cal D} \bar{d}) & \rightarrow & \pi^0 + (e^- + \nu_e) \nonumber \\
& \rightarrow & \pi^0 + (\mu^- + \nu_{\mu}) \nonumber \\
& \rightarrow &  \pi^0 + (\tau^- + \nu_{\tau}) \nonumber \\
\label{calDddecay}
\end{eqnarray}
and variants thereof where $\pi^0$ is replaced by any other non-strange
neutral meson. When $u$ in Eq.(\ref{calDdecay}) is
replaced by $c$ or $t$ which subsequently decay, we arrive
at many other decay channels additional to Eq.(\ref{calDddecay}).

\bigskip

\noindent
Employing instead the ${\cal D}$ decays in Eq.(\ref{calDdecay2})
 implies additional decay modes of $({\cal D} \bar{d})$ meson that include
\begin{eqnarray}
({\cal D} \bar{d}) & \rightarrow & \pi^+ + (e^- + e^-) \nonumber \\
& \rightarrow & \pi^+ + (\mu^- +\mu^-) \nonumber \\
& \rightarrow &  \pi^+ + (\tau^- + \tau^-) \nonumber \\
\label{calDddecay2}
\end{eqnarray}
and variants obtained by flavour replacement.  Eqs.(\ref{calDddecay}) and (\ref{calDddecay2}),
merely illustrate a few of the simplest $({\cal D}\bar{d})$ decays. There are many more.

\bigskip

\noindent
Next we consider the lightest TeV baryons in Table 3 with ${\cal Q} = {\cal D}$.
Using the ${\cal D}$ decays from Eq.(\ref{calDdecay}) we find
for $({\cal D}uu)$ decay
\begin{eqnarray}
({\cal D}uu) & \rightarrow & p + (l_i^- + \nu_i). \nonumber \\
  \label{DuuDecay}
  \end{eqnarray}
together with flavour rearrangements. Here, as in subsequent equations,
$i=e,\mu,\tau$.

\bigskip

\noindent
Alternatively, the ${\cal D}$ decays from Eq.(\ref{calDdecay2})
lead to
\begin{eqnarray}
({\cal D}uu) & \rightarrow & N^{*++} + Y^{- -} \nonumber.  \\
&\rightarrow & p + \pi^+ + (l_i^- + l_i^-). \nonumber.  \\
\label{DuuDecay2}
\end{eqnarray}

\bigskip

\noindent
Looking at the TeV baryon $({\cal D} ud)$ the respective
sets of decays corresponding to Eq.(\ref{calDdecay}) are
\begin{eqnarray}
({\cal D}ud) & \rightarrow & n +(l_i^- + \nu_i) \nonumber \\
\label{DudDecay}
\end{eqnarray}
where only the simplest light baryon is exhibited.

\bigskip

\noindent
Corresponding to ${\cal D}$ decays in Eq.(\ref{calDdecay2})
there are also
\begin{eqnarray}
({\cal D}ud) & \rightarrow & p +(l_i^- + l_i^-) \nonumber \\
\label{DudDecay}
\end{eqnarray}
in the simplest cases.

\bigskip

\noindent
Finally, of the $({\cal D}qq)$ TeV baryons, we write out the
decays for $({\cal D}dd)$, first for the ${\cal D}$ decays
in Eq.(\ref{calDdecay})
\begin{eqnarray}
({\cal D}dd) & \rightarrow & N^{*-} + Y^- \nonumber \\
& \rightarrow & n + \pi^- + (l_i^- + \nu_i). \nonumber \\
\label{DddDecay}
\end{eqnarray}
within flavour variations.

\bigskip

\noindent
With the Eq.(\ref{calDdecay2}) decays of ${\cal D}$ there
are finally decays of the sort
\begin{eqnarray}
({\cal D}dd) & \rightarrow & n +(l_i^- + l_i^-) \nonumber \\
\label{DudDecay}
\end{eqnarray}
again with more possibilities by choosing alternative flavours.

\bigskip

\noindent
We now replace the TeV quark ${\cal D}$ by the next heavier
TeV quark ${\cal S}$ and repeat our study of decays whereupon we shall
encounter the first example of decay not only to the known
quarks but also to a TeV quark.

\newpage

\noindent
The TeV quark ${\cal S}$ has possible decay channels

\begin{eqnarray}
{\cal S} &\rightarrow& d + Y^- \nonumber \\
& \rightarrow & d + (e^- + \nu_e) \nonumber \\
& \rightarrow & d + (\mu^- + \nu_{\mu})   \nonumber \\
& \rightarrow & d + (\tau^- + \nu_{\tau})     \nonumber \\
& \rightarrow &         {\cal D}   + Z'               \nonumber \\
& \rightarrow & d + (e^- + \nu_e) + (e^+ + e^-)  \nonumber \\
& \rightarrow & d + (e^- + \nu_e) + (\mu^+ + \mu^-)  \nonumber \\
& \rightarrow & d + (e^- + \nu_e) + (\tau^+ + \tau^-)  \nonumber \\
& \rightarrow & d + (\mu^- + \nu_{\mu}) + (e^+ + e^-)  \nonumber \\
& \rightarrow & d + (\mu^- + \nu_{\mu}) + (\mu^+ + \mu^-)  \nonumber \\
& \rightarrow & d + (\mu^- + \nu_{\mu}) + (\tau^+ + \tau^-)  \nonumber \\
& \rightarrow & d + (\tau^- + \nu_{\tau}) + (e^+ + e^-) \nonumber \\
& \rightarrow & d + (\tau^- + \nu_{\tau}) + (\mu^+ + \mu^-) \nonumber \\
& \rightarrow & d + (\tau^- + \nu_{\tau}) + (\tau^+ + \tau^-) \nonumber \\
\label{calSdecay}
\end{eqnarray}
where we note the opening up of channels due to 
${\cal S}\rightarrow{\cal D}$ decay.

\bigskip

\noindent
With Eq.(\ref{calSdecay}) in mind, the decays
of the TeV meson $({\cal S} \bar{u})$ include

\begin{eqnarray}
({\cal S} \bar{u}) & \rightarrow & \pi^- + (l_i^- + \nu_i) \nonumber \\
& \rightarrow & \pi^- + (l_i^- + \nu_i) + (l_j^+ + l_j^-)  \nonumber \\
\label{calSudecay}
\end{eqnarray}
where the second line involves a ${\cal D}$
intermediary.

\bigskip

\noindent
An alternative to Eq.(\ref{calSdecay}) is
\begin{eqnarray}
{\cal S} &\rightarrow& u + Y^{- -} \nonumber \\
& \rightarrow & u + (e^- + e^-) \nonumber \\
& \rightarrow & u + (\mu^- + \mu^-) \nonumber \\
& \rightarrow & u + (\tau^- + \tau^-) \nonumber \\
\label{calSdecay2}
\end{eqnarray}
which implies additional decay modes of $({\cal S} \bar{u})$ 
\begin{eqnarray}
({\cal S} \bar{u}) & \rightarrow & \pi^0 + (l_i^- + l_i^-) \nonumber \\
\label{calSudecay2}
\end{eqnarray}
and variants which replace $\pi^0$ by another neutral non-strange meson.
Eqs.(\ref{calSudecay}) and (\ref{calSudecay2}),
illustrate sufficiently $({\cal S}\bar{u})$ decays.

\bigskip

\noindent
Turning to the meson (${\cal S} \bar{d}$), we can use
Eq.(\ref{calSdecay}) to identify its possible decays
\begin{eqnarray}
({\cal S} \bar{d}) & \rightarrow & \pi^0 + (l_i^- + \nu_i) \nonumber \\
\label{calSddecay}
\end{eqnarray}
When $u$ in Eq.(\ref{calSdecay}) is
replaced by $c$ or $t$ which subsequently decay, we arrive
at many other decay channels additional to Eq.(\ref{calSddecay}).

\bigskip

\noindent
Employing instead the ${\cal S}$ decays in Eq.(\ref{calSdecay2})
implies additional decay modes of $({\cal S} \bar{d})$ that include
\begin{eqnarray}
({\cal S} \bar{d}) & \rightarrow & \pi^+ + (l_i^- + l_i^-) \nonumber \\
\label{calSddecay2}
\end{eqnarray}
and variants obtained by flavour replacement.  Eqs.(\ref{calSddecay}) and (\ref{calSddecay2}),
illustrate only a few of the simplest $({\cal S}\bar{d})$ decays. There are many more.

\bigskip

\noindent
Next we consider the lightest TeV baryons in Table 3 with one ${\cal Q} = {\cal S}$.
Using the ${\cal S}$ decays from Eq.(\ref{calSdecay}) we find
for (${\cal S}uu$) decay
\begin{eqnarray}
({\cal S}uu) & \rightarrow & p + (l_i^- + \nu_i). \nonumber \\
  \label{SuuDecay}
  \end{eqnarray}
together with flavour rearrangements.

\bigskip

\noindent
Alternatively, the ${\cal S}$ decays from Eq.(\ref{calSdecay2})
lead to
\begin{eqnarray}
({\cal S}uu) & \rightarrow & N^{*++} + (l_i^- + l_i^-). \nonumber.  \\
& \rightarrow & p + \pi^+ + (l_i^- + l_i^-). \nonumber \\
\label{SuuDecay2}
\end{eqnarray}

\bigskip

\noindent
Looking at the TeV baryon $({\cal S} ud)$ the respective
sets of decays corresponding to Eq.(\ref{calSdecay}) are
\begin{eqnarray}
({\cal S}ud) & \rightarrow & n +(l_i^- + \nu_i) \nonumber \\
\label{SudDecay}
\end{eqnarray}
where only the simplest version is exhibited.

\bigskip

\noindent
Corresponding to the ${\cal S}$ decays in Eq.(\ref{calSdecay2})
there are the decays
\begin{eqnarray}
({\cal S}ud) & \rightarrow & p +(l_i^- + l_i^-) \nonumber \\
\label{SudDecay}
\end{eqnarray}

\bigskip

\noindent
For baryon $({\cal S}dd)$, firstly from the ${\cal S}$ decays
in Eq.(\ref{calSdecay}) we have
\begin{eqnarray}
({\cal S}dd) & \rightarrow & N^{*-} + Y^- \nonumber \\
& \rightarrow & n + \pi^- + (l_i^- + \nu_i). \nonumber \\
\label{SddDecay}
\end{eqnarray}
within flavour variations.

\bigskip

\noindent
Secondly, from the Eq.(\ref{calSdecay2}) decays of ${\cal S}$ there
are baryon decays of the type
\begin{eqnarray}
({\cal S}dd) & \rightarrow & n +(l_i^- + l_i^-) \nonumber \\
\label{SddDecay}
\end{eqnarray}
with more possibilities by choosing alternative flavours.

\bigskip

\noindent
We could continue further to study decays
of all the baryons and mesons in our Tables. 
However, it seems premature
to do so until we
know from experimental data the masses and mixings
of ${\cal D,S,T}$.
We remark only that the type
of lepton cascade which we have deliberately
exhibited explicitly in Eq.(\ref{calSdecay})
becomes a more prevalent possibility as the 
lepton number of the decaying hadron increases.

\bigskip

\noindent
We may expect, by analogy with the top quark,
 that the mass of the ${\cal T}$
quark, although probably below $4.1$ TeV for the
symmetry-breaking reason discussed {\it ut supra}
might be not much below. For example it might exceed $3$ TeV 
whereupon the mass of
a (${\cal TTT}$)  baryon could exceed $9$ TeV. 
Since this baryon has high lepton number, it is pair
produced and such pair production is beyond the reach
of the $14$ TeV LHC.  Its study requires a $100$ TeV collider 
of the type under discussion at CERN
and in China. As a foretaste of the physics,
one notable decay of the $({\cal TTT})$ baryon 
is into [$p+ 4(e^+) + 2(\bar{\nu}_e)$].        

\bigskip

\noindent
At the time of writing, the particles exhibited
in our Tables are conjectural. If the bilepton is established 
as the first new elementary particle discovered
since 2012 \cite{ATLAS,CMS}, the existence of all the additional 
baryons and mesons will become a sharp prediction.

\bigskip

\section*{Acknowlegement}

\noindent
We thank the University of Salento for affiliation and S.L.Glashow for 
supporting a bilepton search at CERN.

\end{document}